# Refining Vision Videos

Kurt Schneider[1], Melanie Busch[1], Oliver Karras[1],
Maximilian Schrapel[2], and Michael Rohs[2]

[1] Software Engineering Group, [2] Human-Computer Interaction Group
Leibniz Universität Hannover, Welfengarten 1, 30167 Hannover, Germany

{kurt.schneider,melanie.busch,oliver.karras}@inf.uni-hannover.de
{maximilian.schrapel,michael.rohs}@hci.uni-hannover.de

**Abstract.** **[Context and motivation]** Complex software-based systems involve several stakeholders, their activities and interactions with the system. Vision videos are used during the early phases of a project to complement textual representations. They visualize previously abstract visions of the product and its use. By creating, elaborating, and discussing vision videos, stakeholders and developers gain an improved shared understanding of how those abstract visions could translate into concrete scenarios and requirements to which individuals can relate. **[Question/problem]** In this paper, we investigate two aspects of refining vision videos: (1) Refining the vision by providing alternative answers to previously open issues about the system to be built. (2) A refined understanding of the camera perspective in vision videos. The impact of using a subjective (or "ego") perspective is compared to the usual third-person perspective. **[Methodology]** We use shopping in rural areas as a real-world application domain for refining vision videos. Both aspects of refining vision videos were investigated in an experiment with 20 participants. **[Contribution]** Subjects made a significant number of additional contributions when they had received not only video or text but also both – even with very short text and short video clips. Subjective video elements were rated as positive. However, there was no significant preference for either subjective or non-subjective videos in general.

## 1 Introduction: Shared Understanding and Vision Videos in RE

When a complex technical or socio-technical system is being conceived, overall visions are developed before software requirements can be specified. In development processes like the V-model (www.iabg.de), *system* requirements and *system* design precede *software* requirements. Changes in business processes, complex interactions, or societal change call for stakeholder participation and discourse. However, it is often difficult to convey the concepts and visions to diverse stakeholders [10]. Due to the large number of available options, building software prototypes for all of them is impossible. Details of their scope and impact are initially unclear.

One of the main challenges in requirements engineering (RE) is to create a shared understanding of the future system among developers and different stakeholder groups [14]. Minutes of stakeholder meetings are usually limited to only one facet of various points of view and a shared vision [12]. Several researchers [8, 11, 19] proposed applying videos in RE due to their communication richness and effectiveness [4]. For example, Brill et al. [6] demonstrate the benefits of using ad-hoc videos compared to textual use cases in order to clarify requirements with stakeholders.

In RE, videos of human-computer interaction were used to document system context [15], product vision [6, 23], or scenarios [22, 28, 29]. They were used as input to a requirements workshop [7], for analyzing usability [22], or for complementing specifications [8, 20]. Fricker et al. [12] proposed to record stakeholder meetings on video as a source of authentic requirements. Many approaches use videos but do not report details about how to produce them [6, 8, 18, 28]. This lack of guidance could be a reason why videos are not yet an established RE documentation practice [17, 21]. In the process of eliciting, refining, and validating requirements with video, we investigated two aspects that may contribute to the benefit of videos: (1) Refining visions by presenting alternatives and (2) refining the camera perspective for better emotional involvement. Refining vision videos can empower elicitation and validation. Creighton et al. [8] proposed a high-tech approach to videos in RE. We follow a different line of research.

> *Affordable Video Approach:* While high-end marketing videos obviously help to convince people, we target *affordable videos* that assist in elicitation and validation of requirements and visions. Hence, creating and refining videos should be affordable with respect to effort, time, and resources. We envision a video-based approach for ambitious requirements engineers in ordinary software development teams.

This paper is structured as follows: Section 2 introduces the example application as a background. In Section 3, we describe the concepts of vision videos in RE and of refining them in particular. Related work is presented in Section 4, before we outline the experiment design (Section 5) and report about results (Section 6). Discussion and threats to validity (Section 7) lead to the conclusion (Section 8).

## 2      Application Example: Shopping in Rural Areas

According to Schneider et al. [25, p. 1], "spatial planning problems are characterized by large and heterogeneous groups of stakeholders, such as municipalities, companies, interest groups, women and men, young people and children". Challenges in spatial planning include shrinking population in rural areas. Mobility options are discussed, and shopping opportunities are related to mobility: How can inhabitants of villages and peripheral areas get access to medical services; how can they buy food and daily supplies if grocery stores close down, and public transportation is missing?

Traditionally, neighborhood help or a grocery bus initiative will be discussed in meetings with citizens. Scheduling and conducting those meetings is difficult and usually reaches only a small portion of citizens and stakeholders. Possibilities to participate are initially high and decrease as more and more aspects are decided. According to the "Paradox of Participation", however, interest in participation tends to be low in the beginning and only rises when there is little left to decide. Therefore, it is desirable to enable and motivate stakeholders to start participating early.

CrowdRE stands for technical approaches to support participation of crowds (of stakeholders, citizens, etc.) in requirements engineering. In [25], we proposed to extend the approach beyond RE towards participating in public discourse. The example application chosen for this paper is a sub-aspect of mobility in rural areas. Shopping without shops seems to call for some kind of ordering online and requires an adequate way of delivering the ordered goods. All variants of ordering and delivery require internet access and sophisticated coordination, which must be provided by software. Long before software can be specified, however, stakeholders should get to see the vision and the variants associated with different proposals.

Shopping in rural areas is a real-world application domain of growing importance. This topic has caught public attention and is discussed in newspapers [1]. Findings from the experiment, therefore, apply to the rural context – and may be applicable to other domains with similar challenges. This is, however, beyond the scope of this paper.

## 3  Concepts to Improve the Use of Vision Videos

As outlined above, vision videos are a good representation for communicating what is proposed, and how it would feel to use it. Following our *Affordable Video Approach*, we intend to solicit feedback, questions, and even objections by affordable self-made video clips in order to start effective discourse early.

**Refinement Process:** Stakeholders should be able to participate in the process of comparing alternatives, selecting, and refining options. As refinement progresses, the discussion with all its proposals and questions and rationale will change its nature: From imagining a vision over defining system alternatives to finally narrowing down on software requirements. Requirements are derived by refining visions.

**Emotion:** Emotional reactions need to be taken seriously. For example, one variant of delivery is frequently discussed in the media: A parcel service deposits parcels in the trunk of their recipients. This asynchronous delivery to a personal space sounds attractive to many. However, when they see how someone opens a trunk with personal items in it, their emotional reaction is sometimes less positive. Video is always concrete. A video confronts stakeholders with possible scenarios that should be considered. Similar to other prototypes, validation of assumptions and elicitation of unexpected reactions merge when watching vision videos.

**Definition and Investigated Scenarios:** The experiment assumes there is a discussion on shopping in a rural area, as described above. At this point, "some kind of internet delivery" is proposed.

> **Definition:** By the term "**vision video refinement**", we refer to the process of replacing gaps, abstract, or vague parts of a vision video by more concrete or detailed video clips (i.e. short parts of the video).

This definition of vision video refinement expands into three scenarios:
1. **Open question**: As long as no proposal has been elaborated, vision videos can show the problem; stakeholders are then asked for their suggestions.
2. **Closed Choice**: Discussion moderators or requirements engineers identify a small number of pre-selected options. They create vision videos to visualize those options in an affordable way. Those videos are distributed and shown to stakeholders, asking them for feedback, such as advantages and disadvantages, newly arising questions, concerns, decisions with rationale.
3. **Refined Video**: After all open questions have been addressed, selected refinements are embedded in the overall video. Gaps and vague parts have been replaced by selected video clips. The resulting refined vision video can be distributed, shown at town hall meetings, or further refined in social media discussions.

The experiment below covers scenarios (1) and (2): Preparing open questions, and selecting from different variants. Scenario (3) was not included in this experiment since it follows the same pattern on the next refinement level. We decided to show all alternatives (A-B-C and 1-2-3) in one video, one after the other (see Fig. 2). Vision videos should not be longer than a few minutes.

### 3.1 Camera Perspectives

Emotional involvement and stimulation of empathy are considered strengths of video [17]. When stakeholders can literally see what an intended solution would mean for them, they are enabled to judge alternative proposals and to participate effectively in the decision-making process. Stakeholder groups face different challenges and may hold different values. Stakeholder should be represented adequately in a video to improve empathy, e.g. by *actors* of their age group and by *subjects* of an experiment. Inspired by research in the HCI community [2, 13], subjective camera perspective may also emphasize identification of stakeholders with actors while watching a video. We illustrate and define core terms for the remainder of this paper.

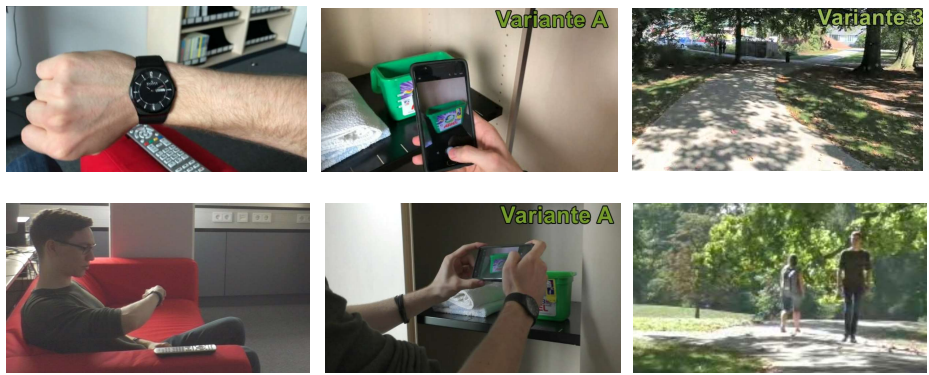

**Fig. 1**. Examples of subjective (top) and corresponding third-person perspective (bottom) from the experiment videos. Variant IDs are displayed temporarily (e.g. "Variante A").

> **Definition: Subjective Camera Perspective**
> In the subjective (also "first-person" or "ego") perspective, a video shows the scene from the perspective of a particular actor. Video seems to be recorded through the eyes of that actor. Audio reflects what the actor hears in that situation.
> **Definition: Third-Person Perspective**
> The situation and scenario is being recorded from an outside point of view. Camera and microphone do not appear (or pretend to be) close to eyes and ears of an actor.

## 4 Related Work

**Vision Videos for RE:** A vision is a positive imagination of the future. It can refer to the capabilities, features, or quality aspects of a part of reality that does not yet exist, but can be imagined. Video is the format in which the vision (content) is presented. Thus, a vision video of a software-based system typically shows a problem, an envisioned solution, and its impact, pretending the solution already exists.

According to this definition, the work by Brill et al. [6] investigated a situation in which one group of subjects created a textual use case specification while a second group prepared a vision video during the same short period of time. Complementary advantages were found. While there was intentional time pressure and inexpensive equipment used in this case, Creighton et al. [8] produced high-end vision videos in cooperation with Siemens and overlaid them visually with UML diagrams. Xu et al. [27] followed this line of research by starting with videos (pixels) and then replacing parts of them with operational software prototypes (bytes). This work demonstrated that visions could successfully be fed into software development activities. In our own work, Karras and Schneider [21] propose developing a quality model for videos that can be used by requirements engineers to produce "good-enough" vision videos. Today, smartphone cameras are of sufficient quality to produce useful vision videos [21]. Practitioners need only a few hints to produce technically sufficient and effective vision videos for eliciting requirements. Pham et al. [23] explored ways of arranging short videos on a custom-made video editor that associated the clips and their arrangement with semantic annotations. Vision videos have been created to promote a societal vision, as in the work of Darby et al. [9] on design fiction: A vision video shows a consultation session of a nurse with a patient. The visionary aspect is a tool that is able to read and interpret body sensors. This tool does not yet exist, but the video pretends it does. The video serves as a visual prototype of the software, its use and context long before even detailed specifications are known. Brill et al. [6] had used videos for the same purpose in our research group. This paper addresses the capability of videos for a discussion process of refinement and discourse rather than for promotional purposes.

**Camera Perspective:** Galinsky et al. [13, p. 110] show how perspective-taking, i.e. "the process of imagining the world from another's vantage point or imagining oneself in another's shoes," decreases stereotyping of others and facilitates social coordination. Aitamurto et al. [2] suspect that the sense of presence may be positively correlated with emotional engagement, empathy, and attitude change as viewers embed themselves in the perspectives of others. The authors suspect that view switching may support taking different perspectives and lead to a better understanding of the perspectives of the different characters, e.g. if the video is filmed in first-person view. Akkil and Isokoski [3] visualized the actor's gaze point in an egocentric video and show that this improves the viewers' awareness of the actor's emotions. Kallinen et al. [16] compared first- and third-person perspectives in computer games and found higher presence for first-person perspective. The concept of *embodiment* in VR refers to the users' experience that the virtual body is perceived as their own. It has been shown that first-person VR environments can create this illusion [26]. This paper analyzes the impact of the subjective perspective in vision videos to refine guidelines for making good vision videos.

## 5 Experiment Design

We used the Goal-Question-Metric Paradigm [5] to formulate goals, hypotheses, questions, and metrics of the experiment.

### 5.1 Goals of Refining Vision Videos

We want to apply vision videos for stimulating discussions on open questions.

> **Main Improvement Goals**: (1) We want to support the process of making choices by refining a vision into more detailed and concrete scenarios. (2) As a separate measurement goal, we want to explore the impact of a subjective camera perspective.

Goal 1 can be rephrased into GQM format: (**Purpose**) Analyze and compare (**Quality Aspect**) number of (new) contributions (**Object**) in feedback (**Perspective**) from young adults. Various combinations of text and video are compared, as specified below.

**Research Questions**: In particular, we are interested in the benefit of providing a second medium. With respect to the GQM goal statement, we investigate whether new contributions can be raised ("stimulated") by video and text, respectively. The camera perspective is directly related to Goal 2 above.

RQ1:  Can adding videos stimulate discussion better than text alone?

RQ2:  Can adding text stimulate discussions better than video alone?

RQ3:  Does a subjective camera perspective in refined vision videos help to empathize with the actor representing a stakeholder?

### 5.2 Video Set-Up and Experiment Procedure

The chosen study design leads to a simple and uniform process of conducting subject sessions. We describe here how the procedure unfolds, and explain our rationale with respect to answering the research questions while considering threats to validity.

**Approach to Refining a Vision Video:** In a live or online discussion on rural shopping, discussions led to identifying ordering and delivery as two crucial open issues. Each subject chooses one refinement A-B-C for ordering, and one refinement 1-2-3 for delivery. Offered options were: (A) Ordering by taking a picture, (B) using a Dash

Button, and (C) a self-ordering sensitive box. Delivery was offered (1) through neighbor-pickup, (2) drones, and (3) deposit in the trunk of a parked car. We used individual sessions for each subject. They saw the videos and texts on a laptop. On the side, they completed the paper questionnaire. Q1 to Q8 are the feedback "object" of Goal 1.

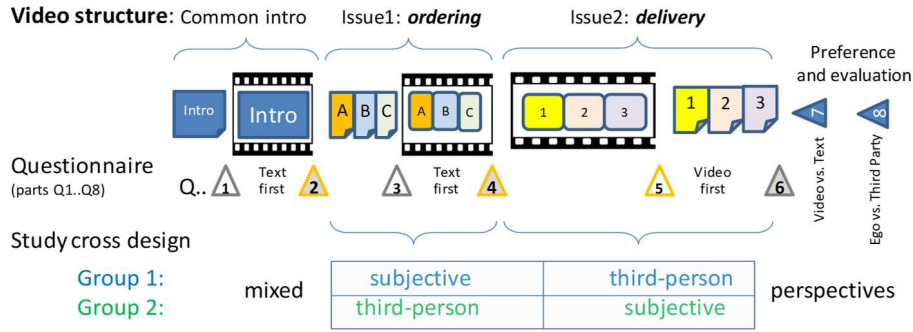

**Fig. 2**: Video structure and experiment design, with Questionnaire parts Q1..Q8

In the experiment, we followed the procedure depicted in Fig. 2. We provided a scenario of buying groceries with two open issues (halt points): (Issue 1) "How can groceries be ordered?" and (2) "How are they delivered?" Subjects completed a questionnaire with eight parts Q1..Q8: Triangles in Fig. 2 indicate what parts of the questionnaire were answered when. For example, Q1 asks for ideas for rural shopping after reading the intro text. Q2 was completed after an introductory video was shown.

There are **two groups of subjects** in the experiment (Fig. 2). Group 1 saw *subjective style videos first* (for ordering), and then third person videos (delivery). Group 2 *started with third-person videos* (ordering), and then saw subjective videos for delivery. This cross design is supposed to mitigate learning effects while at the same time exposing every subject to both camera perspectives. It is important to note that *the presented alternatives (refinements A-B-C and 1-2-3)* of ordering and delivery *must be shown in the same order to all subjects:* They are part of the stimulus that must be kept unchanged in order to produce comparable results.

### 5.3 Hypotheses

In the first block of hypotheses, we investigate subjective aspects of the research questions which are devoted *preference*. The second block of hypotheses investigates the *performance* in terms of the number of contributions. In particular, we investigate the following alternative hypotheses which represent our assumptions. Null hypotheses are reported in the result section (Section 6) together with the results they refer to.

**Preference: What do subjects like?**

Preference is measured by directly asking subjects about their opinion. In the experiment, such a rating is collected after several steps of the experiment.

- H$1_1$: Subjects prefer obtaining a video in addition to a text to getting only the text.
- H$2_1$: Subjects prefer obtaining a text in addition to a video to getting only the video.
- H$3_1$: There is a difference between the Group 1 and Group 2 in how much they like the subjective perspective.
- H$4_1$: Subjects' preference differs between the subjective or third-person perspective.

**Performance: Added contributions to RE and shared understanding**

Performance is measured by counting contributions (GQM "quality aspect"). In the context of RE, *we consider new ideas, new questions, requirements, and rationale as "contributions" for improving shared understanding*. We did not count meaningless and repetitive contributions. The quality of contributions was not rated or evaluated. In this context, the term "*idea*" refers to a contribution about a new form of ordering or delivery.

When information is first represented as text and then a video is added, the benefit of that video is measured in terms of the number of new ideas and new contributions (see above) compared to the ideas respectively the contributions made after seeing only text before. In the inverse case, a video is presented first, and then a text is added:

- H$5_1$: Providing a video in addition to a text leads to new solution ideas.
- H$6_1$: Providing a video in addition to a text leads to new contributions.
- H$7_1$: Providing a text in addition to a video leads to new contributions.

**Emotional effect of the camera perspective: Which of the two perspectives has a greater emotional potential?**

Emotional effect of the camera perspective is measured by directly asking subjects about their opinion. In the experiment, such a rating is collected after subjects saw both types of videos, i.e. in subjective and in third-person perspective.

- H$8_1$: There is a difference in the subjects' perceived emotional involvement of between Group 1 and Group 2.

### 5.4 Selection of Actors, Subjects, and the Affordable Video Approach

There are obviously various age groups of stakeholders affected: Seniors with limited mobility, but also young people on the verge of leaving the village. Seniors and young adults will probably react differently to variants, and they will evaluate them from a different perspective. This important fact is obvious in videos. For this experiment, we focused on the group of young residents. A young actor in a room with modern furniture and big-screen TV has more appeal for empathy to young experiment subjects than a senior in a traditional living room – and vice versa. We collected data (ratings, evaluations, and contributions) from 20 subjects, aged between 20 and 33 years ($M = 25.2$ years). Seven were women, 13 men. We randomly asked members of the intended age group to participate, e.g. during an Open-Door day at the university. Nineteen of them use online shopping, but only eight of them had bought articles of daily use online.

According to our *Affordable Video Approach,* all video clips together were recorded within 3:15 hours of a single day. They were cut using ordinary video editing software within another half day. Video equipment consisted of a standard video camera (300 €) with Rode microphone (330 €) attached, since we found comprehensible audio important in earlier work [25]. Subjective video clips were recorded using a mobile phone camera mounted on a Gimbal (180 €) for the subjective video parts. Mobile phone cameras also would have been sufficient. All four lay actors and video personnel were members of the research group with no advanced video background or training.

The texts for introduction, ordering, and delivery variants are typically read by subjects in silence (32 s for intro, 29 s ordering, and 35 s delivery). Subjective videos on ordering run for 60 s (all three variants together), and 68 s in normal camera perspective. Delivery is more complex and includes interaction beyond the (first person) actor. Delivery variants run for a total of 155 s (subjective) and 150 s (third-person).

## 6   Experiment Results

For evaluating the alternative hypotheses in 5.3, we state corresponding null hypotheses. We provide additional descriptive analysis of ratings, evaluations, and subject opinions as boxplots. Results are clustered in the same three above-mentioned categories: Preference, performance, and emotional effect of the camera perspective.

**Preference**

$H1_0$: Subjects' preference does not differ between obtaining a video in addition to a text and getting only the text.

Subjects had first received a text describing the ordering options and then an additional video illustrating the same ordering options. After watching the video, we asked whether they preferred having the video in addition to the text, or only the text (see Fig. 2, Q4). According to a chi-square test of independence ($\chi^2 = 1.05$, $p = .3$), there is no difference between the two groups. Thus, we could aggregate the data for analysis. Since we had nominal data, we performed a chi-square goodness-of-fit test with a significance level $\alpha = .05$. Corresponding to $H1_0$, one would expect a 0.5/0.5 distribution of the stakeholders' preference. We found significant deviation from the hypothetical distribution ($\chi^2 = 12.8$, $p = .0003$). We can reject $H1_0$ and accept $H1_1$. *Subjects prefer obtaining a video in addition to text rather than having only the text.*

$H2_0$: Subjects' preference does not differ between obtaining a text in addition to a video and getting only the video.

Subjects had first received a video illustrating the delivery options and then an additional text describing the same delivery options. After reading the text, we asked whether they preferred having the text in addition to the video, or only the video (see Fig. 2, Q6). We performed a chi-square test of independence ($\chi^2 = 1.25$, $p = .26$), which indicates no difference between the two groups. Since there is no difference between the groups, we aggregated the nominal data. We found a significant deviation from this distribution ($\chi^2 = 7.2, p = .007$). Thus, we can reject $H2_0$ and conclude: *Subjects prefer obtaining a text in addition to a video rather than having only the video.*

H3$_0$: There is no difference between Group 1 and Group 2 in how much they like the subjective perspective.

At the end of the experiment, the subjects assessed the statement: "I liked the ego-perspective." on a Likert-scale from 0 (totally disagree) to 5 (totally agree) (see Fig. 2, Q8). According to Kolmogorv-Smirnov ($K = .19$, $p = .07$) and Shapiro-Wilk tests ($W = .9$, $p = .05$), the data is normally distributed. Next, we performed a Mann-Whitney U test. The test indicated that the rating of Group 1 ($Mdn = 4$) for the subjective perspective was significantly higher than for Group 2 ($Mdn = 2.5$), $Z = 2.35, p = .02$. Thus, we can reject H3$_0$. *There is a difference between Group 1 and Group 2 in how much they like the subjective perspective.*

H4$_0$: Subjects consider both subjective and third-person perspectives equally good.

We asked subjects if they preferred subjective or third-person perspective (see Fig. 2, Q8). According to the chi-square independence test ($\chi^2 = 1.14$, $p = .56$), there is no difference between the two groups and we can aggregate the nominal data. We applied a chi-square goodness-of-fit test ($\alpha = .05$). According to the $H4_0$, there would be a .5/.5 distribution. We found no significant deviation from the hypothesized distribution ($\chi^2 = 1.125$, $p = .29$). We cannot reject H4$_0$. *There is no significant difference between the subjects' preference for one of the two perspectives.*

We asked how emotionally involved subjects were after seeing the second medium (video after text / text after video). Fig. 3 (left) shows the high ratings on a 0 to 5 Likert scale. In all three cases (introduction, ordering, delivery) the emotional involvement was higher after receiving the second medium. All videos received very high ratings ($Mdn = 4$); the stimulated emotional involvement. With text, values are a little lower.

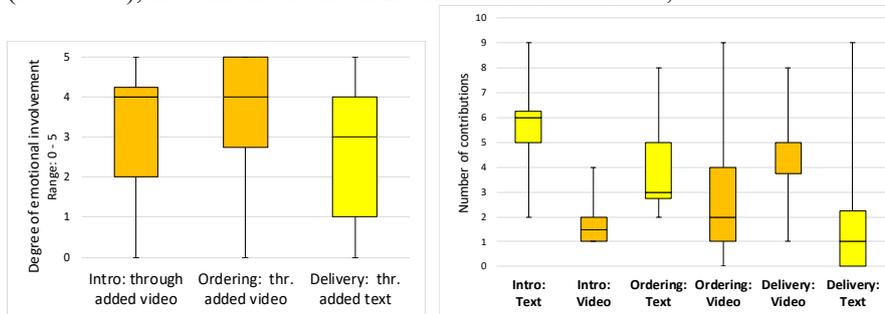

**Fig. 3**. Emotional involvement at Q2, Q4, Q6; No. of contributions at Q1/2; Q3/4; Q5/6

**Performance**

The performance is measured in number of contributions after text or videos were provided. Fig. 3 (right) shows the three parts: introduction, ordering, and delivery. Boxplots on the left of each pair show the number of contributions made after the first element was provided; the right-hand boxplots show additional contributions solicited after the *second* element was provided. Light boxes stand for text, darker ones for video.

H5$_0$: Providing a video in addition to a text does not lead to new solution ideas.

Subjects had first received text and were asked to write down solution ideas (see Fig. 2, Q1). After participants had received the video, we asked if they had any additional solution ideas (see Fig. 2, Q2). According to Kolmogorov-Smirnov ($K = .34$, $p < .001$) and Shapiro-Wilk ($W = .81$, $p = .001$) tests, the number of solution ideas are not normally distributed. We investigated whether the two groups differ from each other by using Mann-Whitney U test: $Z = .53$, $p = .60$. Since we did not find a difference between the two groups, we aggregated the data and performed the non-parametric one-sample Wilcoxon Signed-Rank test. The test showed a significant number of additional, not yet mentioned, solution ideas by the stakeholders ($Z = -3.62$, $p < .001$). $H5_0$ is rejected: *Providing a video in addition to text leads to new solution ideas*.

$H6_0$: Providing a video in addition to a text does not lead to new contributions

After the subjects read the text of the ordering options we asked them to select one option and to write down their rationale, requirements, and questions (Fig. 2, Q3). Afterwards the participants received a video of the ordering options and we asked them for further requirements-related contributions (Fig. 2, Q4). We investigated the collected number of contributions for normal distribution with Kolmogorov-Smirnov test and Shapiro-Wilk test. Both tests indicated that the data is not normally distributed ($K = .21$, $p = .02$, $W = .89$, $p = .02$). There is no difference between the groups by means of a Mann-Whitney U test: $Z = .91$, $p = .36$. We analyzed all data together by using the one-sample Wilcoxon Signed-Rank test. This test yields a significant difference, i.e. a significant number of new contributions ($Z = -3.62$, $p = .0002$). $H6_0$ is rejected: *Providing a video in addition to a text leads to new contributions*.

$H7_0$: Providing a text in addition to a video does not lead to new contributions.

For the delivery options, subjects saw the video first and we asked them to select one option. Based on their choice, we asked them to write down their rationale, requirements, and questions (Fig. 2, Q5). Then they read the text describing the delivery options and we asked them for further requirements-related contributions (Fig. 2, Q6). The statistical analysis follows the same procedure: The Kolmogorov-Smirnov test ($K = .27$, $p < .001$) and Shapiro-Wilk test ($W = .74$, $p < .001$) showed that the data is not normally distributed. There was no difference between the groups in a Mann-Whitney U test: $Z = .76$, $p = .45$. We analyzed all data together by using the non-parametric one-sample Wilcoxon Signed-Rank. The test yields a significant number of additional contributions after the subjects read the text ($Z = -3.06$, $p = .001$). $H7_0$ is rejected. *Providing a text in addition to a video leads to new contributions*.

**Emotional effect of the camera perspective**

$H8_0$: There is no difference in the subjects' perceived emotional involvement between Group 1 and Group 2.

Subjects first received the text of the variants and then the video. Afterwards, we asked subjects to indicate their emotional involvement by assessing the statement "I was more emotionally involved in the problem due to the video." on a Likert-scale from 0 (totally disagree) to 5 (totally agree) (see Fig. 2, Q4). While the Kolmogorov-Smirnov test ($K = .20$, $p = .07$) indicated that the data is normally distributed, the Shapiro-Wilk test found the data to be not normally distributed ($W = .88$, $p = .03$). Due to this discrepancy, we used a non-parametric Mann-Whitney U test. It showed no difference between the group that watched a subjective video ($Mdn = 4$) and the group that watched a third-person video ($Mdn = 3$), $Z = .44, p = .66$. We cannot reject H8$_0$ and conclude: *There seems to be no difference between Group1 and Group 2 in the perceived emotional involvement of subjects.*

**Evaluations and Subject Opinions**

Finally, we asked subjects in Q7 for more detailed feedback (Fig. 4) after all texts and videos had been presented.

a) Text important for choosing a variant
b) Video important for choosing a variant
c) I liked the videos
d) Videos provide important information
e) Videos convey atmosphere
f) Video quality was sufficient
g) Videos were obsolete

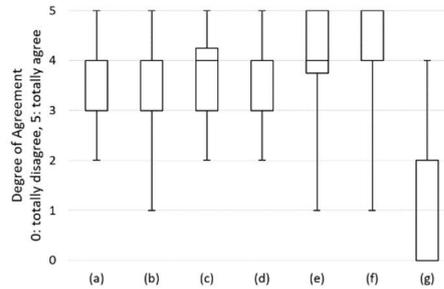

**Fig. 4.** Subjects' detailed evaluation results. (g) means: "videos were **not** obsolete"

As Fig. 4 indicates, both text (a) and video (b) were considered important for making decisions about the refinement variants. Most subjects liked the videos (c). Most subjects also found videos provided important information (d). The ratings for "video conveys atmosphere" (e) were even higher, but there were also a few low ratings. In (f), a large majority considered video quality sufficient, despite the *Affordable Video Approach*. Most disagreed with "videos were obsolete" (g) - not an obvious result, given the short runtime of all videos and the fact that there were also texts available.

## 7   Interpretation and Discussion

We investigated whether adding videos to previously available texts would solicit additional contributions for discourse. In Q6, we also measured the inverse situation: Adding text to previously shown videos. In real decision situations about rural areas, most stakeholders would read brief texts (in the newspaper [1], or online) before they decide to watch a short video about it. The results confirm the usefulness of enriching early discussions about visions and requirements with both text and video. Preference and evaluation were very positive, and a number of statistically significant results confirm that adding either video or text (to the other) stimulated more contributions.

### 7.1   Threats to validity

The experiment design presented in Sec. 5 is rather sophisticated. It reflects the complexity of evaluating the role of vision videos in refining visions towards requirements. The real-world application domain is of substantial complexity. Hence, a number of specific threats to validity must be considered.

**Internal validity**: *Causality.* Possible influences on the outcome are numerous, due to the high complexity. For the experiment, we used texts and videos that were created independently. Neither did one build on the other, nor was there a competition to outperform each other. The mission was to explain the introduction and refinement options concisely and self-sufficiently. Some subjects might have felt pressed to provide more contributions when they were shown an extra video or text. However, we checked whether those new contributions were original or repetitive and counted only new ones. There were several cases in which a question did not solicit any additional responses.

**External validity**: *Can results be generalized?* Despite the above-mentioned precautions, our findings cannot be generalized directly to every kind of text and every type of video. Texts and videos can be useful or difficult to understand and annoying - intentionally or by accident. There are so many types and styles of video (and text) that one should take even our significant findings with a grain of salt.

**Construct validity**: *Adequate concepts?* As explained in Sec. 5, we counted new questions, new reasons to choose or reject a variant as *contributions* to the discourse as RE contributions. In the area of RE, a good question can be valuable [14] for clarifying visions and requirements. The results and findings should be read with this definition in mind. Conceptualizations of "contribution" that deviate substantially from this definition may not be covered by our experiment. The treatments in our experiment was adding a second medium (video/text). We analyzed the effect of getting that treatment by comparing contributions before and after receiving the second medium.

**Conclusion validity**: *Adequate conclusions?* The positive impact of adding video or text could be a case of "paraphrasing", presenting a situation from different angles. It is possible and likely that adding other media could have had similar effects. We wanted to investigate whether low-effort video with its own advantages was also qualified as a useful second medium. *Our results confirm the benefit of taking the extra effort of providing a second medium.* Please note that providing "both video and text at a time" may seem a similar and attractive option, but poses yet another threat to validity: Its impact will depend highly on how the media are presented: if and in which order subjects look at them. This aspect was beyond the scope of our study.

Deciding about rural shopping is an integral part of much wider concerns. There are so many parameters and influence factors that cannot – and should not – be controlled in order not to distort the phenomenon of interest. We decided to study the very basic mechanisms of refining a vague vision of shopping into several variants by video. The technical choice of a camera perspective is related. Those mechanisms are the basis for more complex interactions of vision and requirements communication and clarification.

## 8 Conclusions

We had expected to stimulate additional questions and ideas by showing videos – where usually only a few short texts would be provided. However, we did not expect the number of additional contributions stimulated by the videos (Fig. 3), and the very positive evaluation of videos in hindsight (Fig. 4). In the introduction to the experiment, exactly the same text was provided to *read* – and then to *hear* in the video. Nevertheless, almost all subjects recommended showing the video in addition to the text in the future. We had included the inverse direction (text after video) in the study as a matter of curiosity. Given the less than 10-line texts and 20-second video clips describing an ordering refinement, we had not expected performance and preference indicators to be as clear as they were: *Provide both media*.

Subjective camera perspective seemed to be a matter of taste. There was no significant performance advantage over third-person videos, nor was the empathy rating higher. Some subjects preferred subjective over third-person perspective – and vice versa. According to Runeson et al. [24], case studies are appropriate for investigating phenomena that require complex subsets of reality to occur. Based on this established baseline of experimental insights, we plan to triangulate our findings in case studies.

## Acknowledgement

This work was supported by the Deutsche Forschungsgemeinschaft (DFG) under Grant No.: 289386339, project ViViReq. (2017–2019).